# Understanding interface properties in 2D heterostructure FETs

Kosuke Nagashio*
*Department of Materials Engineering, The University of Tokyo, Tokyo 113-8656, Japan*
*nagashio@material.t.u-tokyo.ac.jp

**Abstract**
Fifteen years have passed since graphene was first isolated on the substrate from bulk graphite. During that period, 2D layered materials with intrinsic band gaps have been realized. Although many exciting results have been reported for both their fundamental physics and applications, the discussion of 2D electron device application to the future integrated circuit is still based on the expectation of the inherently high properties that 2D materials ideally possess. This review article focuses on the gate stack property, which is one of most important building blocks in the field effect transistor. Starting from the comparison of the 2D/SiO$_2$ interface properties with the conventional SiO$_2$/Si interface properties, recent advances in the studies of gate stack properties for bilayer graphene and MoS$_2$ FETs are discussed. In particular, the advantages and disadvantages of the 2D heterostructures with 2D insulator of *h*-BN are emphasized. This review may provide conceptual and experimental approaches for controlling the 2D heterointerface properties.

**1. Introduction**

The realization of a new channel device field-effect transistor (FET) that achieves both ultralow power consumption and high-speed operation is indeed difficult in the integrated circuit that forms the basis of next-generation electronics, but there can be no doubt regarding its necessity. For MoS$_2$, which is a typical two-dimensional layered material, the recent demonstration of an effective channel length of ~3.9 nm in FET [1] and a quite low subthreshold swing (S.S.) of ~3.9 mV/dec in tunnel FET [2] has facilitated research on 2-dimensional (2D) layered channels. Although many outcomes have already been obtained regarding the fundamental physical properties and device applications through the highly competitive worldwide research effort, the prospects of 2D electronic device applications are still discussed based on the potentially high properties of the materials themselves [3]. This is largely related to the following two reasons. First, the high-quality wafer-scale synthesis of 2D materials has not been commercialized yet, even though it has been demonstrated at the academic research level [4,5]. Second, a common strategy for controlling 2D channels, even in a single transistor structure, has not been established yet.

Looking back on the past electronic and optical devices, one of the historical breakthroughs was the understanding and control of the gate stack, which is one of most important building blocks in the field effect transistor, more specifically, the SiO$_2$/Si interface

**2. Comparison of the SiO$_2$/Si interface and 2D vdW interfaces**
**2.1 Advantage of 2D layered channels against Si**

The metal-oxide-semiconductor field-effect transistor (MOSFET) scaling requires the transition properties and the lattice mismatch at the heterointerfaces in compound semiconductors. Compared with conventional 3D semiconductors, the stacking of various layers in the 2D heterostructures is possible without considering the lattice mismatch due to the van der Waals (vdW) interface. 2D layered heterostructures provide a new scheme for device applications because an electrically inert interface is "ideally" expected. However, detailed studies of the 2D heterostructure interface properties have been quite limited [6-14] because it is difficult to apply conventional C-V measurements to 2D systems. Therefore, we have faced this issue squarely in our work to date [15-21]. In this review article, the SiO$_2$/Si interface properties revealed by many dedicated researchers are first compared with the fundamental properties of graphene and MoS$_2$ on the substrates. Then, the recent progress in the studies of 2D heterostructure interface properties for bilayer graphene and MoS$_2$ is described. For the bilayer graphene/*h*-BN heterostructure FET, an electrically inert interface and spatially uniform band gap opening can be demonstrated. For MoS$_2$ FET, full energy spectra of the interface states densities were obtained and their origins was understood by comparing the data for various kinds of gate stack structures. The purpose of this review is to describe the conceptual and experimental approaches for controlling 2D heterointerface properties.

from planar to FinFET to overcome the short channel effect, prolonging the life of the silicon complimentary MOS (Si CMOS) [22,23]. Nevertheless, at the 2015 IEEE International Electron Devices Meeting (IEDM), Dr. Yeric of ARM research has stated that an attractive charge-based device scaling path would be back to 2D from 3D [24]. The reason behind this opinion can be understood in **Fig. 1(a)**. While electrostatics have been improved by the FinFET structure, the 3D shape has drastically increased the parasitic capacitance ($C_{para}$) to a level beyond the intrinsic gate capacitance, resulting in the loss of gate controllability [25]. In Dr. Yeric's plenary lecture, a high-performance race car with a large parachute was used as a metaphor, as shown in **Fig. 1(b)**.

Here, what is the difference between the conventional 2D channel of Si, i.e., silicon-on-insulator (SOI), and the 2D layered channels? Since the 2D channel is expected to be used complimentarily with advanced Si-CMOS, the monolayer channel with the short channel length less than ~10 nm must be considered. If the 2D channel thickness is ~10 nm, it is not 2D anymore. Therefore, here we focus on the monolayer 2D channel. **Figure 2** compares the square modulus of the wave functions of the subbands and the gate-channel capacitance ($C_{gc}$) for (a) 5-nm SOI and (b) monolayer 2D. The carriers confined in the electric-field-induced band bending are quantized into a 2D subband structure, where $E_1$ and $E_2$ correspond to the two lowest energy subbands. The distance between the centroid for the square modulus of the wave function and the interface is shown by $z_{inv}$. Therefore, the capacitive contribution resulting from this distance in the semiconductor should be added to $C_{gc}$, resulting in $1/C_{gc} = 1/C_{ox} + 1/C_{inv}$, where $C_{ox}$ is the geometrical oxide capacitance and $C_{inv}$ is the inversion layer capacitance. By reducing the thickness of SOI from 25 nm to 7 nm, the quantized carriers are further confined by the channel thickness, resulting in the sharper carrier distribution and slightly enhanced $C_{gc}$ [26]. On the other hand, for $MoS_2$, the drastic increase in $C_{gc}$ at the accumulation is clearly observed in the monolayer limit [20]. Because the monolayer thickness is 0.65 nm, the electron density exists just at the interface, thereby neglecting $z_a^{2D}$. That is, $C_{gc}$ can be approximated as $\sim C_{ox}$. Thus, it is understood that the gate controllability in 2D is superior to that of 7-nm SOI. Moreover, the largest difference between SOI and a 2D layered channel is found for the carrier mobility. When the Si thickness decreases from 5 nm to 1 nm, a drastic mobility degradation is observed, and the fabrication of an SOI with a uniform thickness of 1 nm is extremely difficult [27-29]. On the other hand, the mobility for the monolayer 2D layered channels can be retained and can be as high as several hundreds of cm$^2$/Vs at room temperature, because all of the bondings are closed within the layer [30,31]. Therefore, the superiority of atomically thin 2D channels can be understood from the perspective of electrostatics, $C_{para}$ and carrier mobility at the monolayer limit.

### 2.2 Interface control

The electric field effect enables the modulation of the carrier density in a semiconductor channel. To date, energy-efficient transistor operation at *S.S.* = 60 mV/dec has been achieved by reducing the interface states density ($D_{it}$) at the $SiO_2$/Si interface. Because the insulator/2D layered channel interface possesses a different atomic structure from the $SiO_2$/Si interface because it is a vdW interface, a detailed understanding of the 2D layered channel interface is required to further improve the device performance.

**Figure 3** summarizes the characteristics of $SiO_2$/Si, graphene, and $MoS_2$. For the thermally oxidized $SiO_2$/Si system, the suboxide region with a thickness of ~0.2 nm is formed due to the change in the chemical composition at the interface. Moreover, because the Si electronic states are composed of $sp^3$ hybrid orbitals, the Young's modulus of ~200 GPa for Si is larger than that for amorphous $SiO_2$. The compressive stress formed by the difference in their volume expansion coefficients during the thermal oxidation is predominantly introduced on the $SiO_2$ side with a thickness of ~1 nm [32,33]. This transition region in $SiO_2$ has been clearly recognized by the softening of LO phonon for $SiO_2$ [34]. Although a structural transition region with a thickness of only 3 atomic layers is known to exist on the Si side [35,36], the Si inside this thin transition region possesses the ideal structure. Therefore, the interface is known to be chemically quite steep. For a 5-nm SOI, as shown in **Fig. 2**, the maximum carrier density position is located slightly away from the $SiO_2$/Si interface, enabling the mobility in SOI to be maintained from the perspective of the carrier scattering. These structural defects at the $SiO_2$/Si interface, i.e., the Si dangling bonds ($P_b$ centers), and the spatial fluctuations of the Si-Si bonding angle are known to be introduced as the interface states [37,38]. The advancement of the Si process technology has enabled the control and reduction of $D_{it}$, establishing the status as the basic element in the present integrated circuit.

By contrast, in monolayer graphene, the $sp^2$ hybrid orbitals constitute the honeycomb lattice, while $\pi$ electrons are responsible for the carrier transport. Although these $\pi$ electrons possess the conventional parabolic energy dispersion, linear dispersion appears near the Fermi level ($E_F$) because the half filling condition in which one electron occupies one atom is satisfied, and $E_F$ stays just at the intersection of the bonding band and antibonding band (i.e., the Dirac point). Such linear dispersion give rise to quite unique physical phenomena that have never been realized in conventional semiconductors [39]. Monolayer graphene transferred on the $SiO_2$/Si substrate partially follows the surface topography of the $SiO_2$ substrate (RMS = ~0.2 nm). Because the Young's modulus of graphene (1 TPa) is highest among known materials [40], the introduced strain does not affect the electronic dispersion, and the local linear dispersion can be clearly retained. Meanwhile, because the density of states (DOS) near the Dirac point is quite small for graphene, the Dirac point in graphene spatially fluctuates relative to $E_F$ due to the charged impurities that exist near the SiO$_2$ surface. This spatial fluctuation is in the energy range of ~20-50 meV as the average for the entire channel [41,42].

Finally, MoS$_2$ is a covalent bonding system, and the electronic states are formed by the ligand field [43,44]. Four electrons are transferred from Mo (electronic configuration: Kr4$d^4$5$s^2$) to S (electronic configuration: Ne2$s^2$3$p^4$), and six-coordinated Mo formed by the positively charged $d^2sp^3$-hybridization is placed in the ligand field formed by the negatively charged $sp^3$-hybridized S. In this case, the Mo $d$ orbital is split through the interaction with the ligand orbital. The unique characteristic of MoS$_2$ is that the split $d$ orbitals constitute both the conduction and valence bands, as shown in **Fig. 3**. The band structure formed by this ligand field can be expected to be easily affected by the external strain due to the relatively small Young modulus of MoS$_2$. In fact, scanning tunneling microscopy (STM) measurements of the band gap ($E_g$) of monolayer MoS$_2$ transferred on the SiO$_2$ substrate found large spatial fluctuations, particularly for the valence band side [45].

Here, the defect densities for graphene, $h$-BN, and MoS$_2$ are compared in terms of crystallinity of the 2D bulk crystals. For Kish graphite, it is generally difficult to detect the defects by STM, suggesting that the defect density is negligibly small. For $h$-BN grown at high temperature and high pressure by Taniguchi at the National Institute for Materials Science (NIMS) [46], a defect density of ~$10^9$-$10^{10}$ cm$^{-2}$ has been reported by STM [47]. For MoS$_2$, different defect densities have been reported for the samples obtained by different growth methods. For the MoS$_2$ crystal grown by the chemical vapor transport (CVT) method and MoS$_2$ obtained by the chemical vapor deposition (CVD) method, S vacancies (V$_S$) of ~$10^{13}$ cm$^{-2}$ have been repeatedly detected, while anti-site defects have been reported for MoS$_2$ obtained by physical vapor deposition (PVD) [48,49]. Because the defect densities for graphite and $h$-BN are much smaller than the typical $D_{it}$ value detected at the conventional SiO$_2$/Si interface, external factors are likely the origin of $D_{it}$ in the graphene and $h$-BN system. For MoS$_2$, because the defect states cannot be excluded as an origin for $D_{it}$, both external and internal factors should be considered.

As mentioned above, in the case of a 3D Si system, the strain introduced by thermal oxidation is basically relaxed in the entire SiO$_2$ side, and physical and chemical discontinuities related to the interface structures contribute to the interface states. By contrast, for the 2D systems on the SiO$_2$/Si substrate, the dominant external factors are different, i.e., the spatial fluctuation of charged impurities for graphene and physical strain for MoS$_2$. However, these external factors do in fact arise from the SiO$_2$/Si substrate. When these are transferred on the $h$-BN substrate with an atomically flat surface and negligible charged impurities, the spatial fluctuation of the Dirac point in graphene and the spatial fluctuation of $E_g$ in MoS$_2$ are strongly suppressed [50,45]. From the viewpoint of the miniaturized FETs in the integrated circuit, when the channel size becomes comparable to the size of the spatial fluctuation of the potential, the effect on the device operation is very strong. Therefore, the 2D/$h$-BN heterostructure is quite promising. Nevertheless, the conventional high-k/2D is still quite important because the dielectric constant of $h$-BN ($k$ = ~3) is smaller than that of SiO$_2$. Although the understanding of the external factors for the interface properties in the 2D systems has been obtained continuously from local analysis using STM, detailed research on the interface properties for both high-$k$/2D and 2D/2D interfaces is still quite limited for the analysis of FET devices. In this review article, our recent results on interface-related research are described.

## 3. All 2D heterostructure BLG FET

Monolayer graphene with linear electron dispersion has no band gap, while bilayer graphene (BLG) possesses an electrostatically tunable but relatively small $E_g$ of up to ~0.3 eV attained by applying a displacement field ($\bar{D}$) [51]. The displacement field induces different carrier densities in the top and bottom layers of BLG, giving rise to a potential difference between the two layers. This inversion symmetry breaking is the physical origin behind the band gap opening [52]. Although $E_g$ of ~250 meV has been detected optically [53], so-called "gap states" exist in BLG, and the nearest neighbor hopping transport results in the high off-current ($I_{off}$) [51,54]. Here, to understand the origin of the gap states and the interface properties for BLG, two different gate stack structures are systematically discussed.

The BLG FET with high-$k$ Y$_2$O$_3$ top gate was fabricated on a SiO$_2$(90 nm)/$n^+$-Si substrate through mechanical exfoliation from Kish graphite, as shown in **Fig. 4(a)**. Generally, the high-$k$ oxide deposited on 2D layered materials by atomic layer deposition (ALD) exhibits a relatively low dielectric constant and density [55]. In this study, the thermally deposited Y$_2$O$_3$ was reoxidized in O$_2$ at 100 atm and the relatively low temperature of 300°C so that no defects are introduced. A high electrical field that was as high as ~7 V/cm and a negligible leak current without hysteresis have been achieved for a thickness of 6 nm [56]. This gate stack is hereafter referred to as "high-$k$/BLG." In this case, a maximum $E_g$ of ~250 meV was achieved, while the current on/off ratio ($I_{on}/I_{off}$) was still three orders of magnitude, even at 20 K [15]. Nevertheless, this $I_{on}/I_{off}$ is the highest reported in the studies on BLG with a

high-$k$ gate stack. Moreover, $D_{it}$ has been detected as ~$10^{13}$ cm$^{-2}$eV$^{-1}$ through analysis of the frequency dispersion of $C_{gc}$; this value is quite high compared to that of the conventional SiO$_2$/Si system. Although many different oxidation methods have been applied to improve the Y$_2$O$_3$/BLG interface, $I_{on}/I_{off}$ was not improved because a reduction of $D_{it}$ was not observed. As discussed in **section 2.2**, the potential fluctuation caused by the charged impurities in both high-$k$ Y$_2$O$_3$ and SiO$_2$ result in the spatial fluctuation of $E_g$, as shown in **Fig. 6(c)**. This band gap fluctuation can be macroscopically considered as band edge disorder that acts as gap states. This is the reason why $I_{on}/I_{off}$ was so small despite the high crystallinity of BLG and Y$_2$O$_3$.

To reduce the potential fluctuation in the gate stack, a micromanipulator alignment system for the fabrication of the vdW hetero gate stack with micrometer size has been constructed, and the conditions for bubble-free pinpoint transfer have been studied in detail [57,58]. Then, an all-2D-heterostructure BLG FET with the graphite/$h$-BN/BLG/$h$-BN/graphite gate stack was fabricated, as shown in **Fig. 4(b)**. In this heterostructure, the graphite flake is used even for the top gate electrode. The Ni electrode deposited on top gate $h$-BN is polycrystalline and consists of tiny grains, leading to an increase in the potential fluctuation in BLG due to the variation of the work function for different crystal orientations [16]. Therefore, the top graphite electrode is highly advantageous. This gate stack is hereafter referred to as "all 2D hetero." In this case, the transfer characteristics are much sharper than those in high-$k$/BLG, as shown in **Fig. 5(b)**. The drastic suppression of $I_{off}$ even at a smaller band gap of ~100 meV is clearly observed at 20 K, and the carrier mobility is ~20,000 cm$^2$/Vs. Because the channel resistance reached ~5 G$\Omega$, $I_{off}$ has already reached the measurement limitation. Therefore, it is not meaningless to discuss $I_{on}/I_{off}$, but it reaches ~5×10$^5$ [16]. Moreover, for $C_{gc}$ in all 2D hetero, it is demonstrated that the electron trap/detrap response at this heterointerface is suppressed to an undetectable level in the measurement frequency range between 1 kHz and 2 MHz.

The charge neutrality point at $\overline{D} = 0$ is shown by arrows in **Fig. 5(a) and (b)**. BLG is quite sensitive to the surroundings. Since the charge neutrality point in **Fig. 5(a)** is shifted to the positive top gate voltage side, the positive charge transfer to BLG from the high-$k$ insulator can be realized. On the other hand, there is almost no shift in the charge neutrality point in **Fig. 5(b)**, indicating no charge transfer to BLG from $h$-BN, which is consistent with the fewer charged impurities present in $h$-BN. From the viewpoint of device operation, the charge neutrality points at both positively and negatively maximum $\overline{D}$, which are shown by the bold arrows in **Fig. 5(b)**, should be adjusted to zero top gate voltage to achieve both $n$-type and $p$-type operation.

This adjustment is generally carried out by properly selecting the work function of the gate metal. For all 2D hetero, this may be difficult since the selection of 2D metals is quite limited. Therefore, other methods are required to adjust the charge neutrality point.

**Figure 6(a)** shows $E_g$ as a function of $\overline{D}$, where $E_g$ was obtained from the temperature dependence of electron transport and $\overline{D}$ is the external electrical field ($D/\varepsilon_0$). For high-$k$ BLG, the larger electrical field can be applied due to the high dielectric constant of Y$_2$O$_3$, and the $E_g$ value reaches the almost maximum value of ~250 meV. Alternately, the $E_g$ value for all 2D hetero is limited to ~100 meV because the maximum electrical field applied by $h$-BN without leakage is ~1.5 V/nm due to the small $k$ (~3) of $h$-BN. Nevertheless, it should be emphasized that $I_{on}/I_{off}$ for all 2D hetero is larger than that for high-$k$/BLG. Here, interestingly, as shown in **Fig. 6(a)**, $E_g$ exists for the high-$k$/BLG case even at $\overline{D}$ = 0 V/nm, suggesting that $E_g$ is formed due to the potential fluctuation of the surroundings ($\sigma_s$ = ~26 meV). Therefore, the potential fluctuation energy is only ~1 meV for all 2D hetero. **Figure 6(b)** suggests that the potential fluctuation energy of ~20-200 meV is included in the conventional semiconductor systems as long as amorphous oxides are used as gate insulators. The effect of the potential fluctuation due to amorphous oxide insulators is more prominent in small-gap semiconductors and topological insulators [59]. Therefore, the vdW layered heterostructure is quite promising for studying the physical properties with low energy (a few meV). Indeed, using this heterostructure with $h$-BN, unconventional superconductivity has been realized in a two-dimensional superlattice created by stacking two sheets of graphene that are twisted relative to each other by a small angle of 1.1° [60].

To summarize, both BLG and $h$-BN are layered materials with closed bonding and negligible defect density. By constructing their heterostructure gate stack, an electrically inert interface and spatially uniform $E_g$ can be realized. From the viewpoint of the miniaturized FETs in the integrated circuit, the reduction of the spatial fluctuation of the potential in all 2D hetero drastically improves the gate controllability. If the spatial fluctuation of the potential is large, a larger gate bias is required to completely cut off the current. This is critical for the miniaturized FETs in which the driving voltage should be reduced. Moreover, $h$-BN possess an atomically flat surface and fewer charged impurities than SiO$_2$. Therefore, the surface roughness scattering and Coulomb scattering can also be dramatically reduced, resulting in the enhancement of the mobility. However, as shown in **Figure 6(a)**, the maximum $E_g$ could not be opened by the $h$-BN insulator because of its small dielectric constant, suggesting that the gate stack structure must be reconsidered. One possibility is the structure such as the combination of h-BN and high-k oxide. Because $E_g$ for BLG can be opened by the carrier density difference between top and bottom layers, it is quite sensitive to the existence of the spatially fluctuating charges in surrounding oxides. To open $E_g$ further while retaining the BLG/h-BN interface quality of the present system, one possible approach is the utilization of high-k single-crystal nanosheets [61] in which potential fluctuation may be reduced due to the crystal periodicity.

### 4. Top gate MoS$_2$ FET

In MoS$_2$ FET operation, the following quite naive idea has become widely accepted despite the lack of direct experimental verification: "High electron doping takes place in the MoS$_2$ crystal due to intrinsic S vacancies. Therefore, MoS$_2$ FET is normally in the on state in FET operation. Then, these S vacancies also act as fixed ions, resulting in the off state by the full depletion of the atomically thin MoS$_2$ film when the negative gate bias is applied." Here, the issue of contact is neglected since the electron is easily injected into the MoS$_2$ channel from most of metals due to Fermi level pinning [62]. This electron doping from S vacancies is considerably different from that in Si where the channel polarity is controlled externally by the substitutional doping. To date, the channel formation behavior in the inversion layer has been widely studied in detail based on the change in $C_{gc}$ in the Si MOSFET [63,64]. Therefore, this ambiguity in MoS$_2$ FET operation is largely due to the limited study of the C-V measurement in the MoS$_2$ FET due to some difficulties. Indeed, in the multilayer MoS$_2$ for which the interlayers are held together by the vdW force, the following questions have not been elucidated. (i) How does the 2D channel capacitance contribute in $C_{gc}$? (ii) Do the depletion layer and inversion layer truly exist? (iii) How can $D_{it}$ be extracted? In this section, our recent analysis of the $C_{gc}$ measurement from the depletion to the inversion in the Al$_2$O$_3$ top gate MoS$_2$ FET is explained [17-21]. Moreover, the advantages and disadvantages of the C-V measurements in MoS$_2$ FET are clarified, and the method for extracting $D_{it}$ from I-V analysis is presented based on the disadvantage of the difficulty of $D_{it}$ extraction from the conventional frequency dispersion in C-V measurements [18,19].

### 4.1 Gate channel capacitance in monolayer MoS$_2$ FET: Quantum capacitance

As discussed in **section 2.1**, the capacitance of the channel itself, as well as $C_{ox}$, is included in the $C_{gc}$ of the SiO$_2$/Si gate stack. When the Si channel is depleted by applying the gate bias, the carriers are extruded to the edge of the depletion layer, and their carriers respond with the alternative excitation voltage during the C-V measurement. Therefore, the depletion capacitance ($C_D$), which is expressed by the depletion layer width ($W_D$) and dielectric constant of Si ($k_{Si}$) as $C_D$ = $k_{Si}/W_D$, is contributed in series with $C_{ox}$ in $C_{gc}$, i.e., $1/C_{gc}$ = $1/C_{ox}$ + $1/C_D$. When the gate bias is further increased, the inversion layer is formed. If the carriers in the inversion layer respond with the alternative excitation voltage, the inversion capacitance ($1/C_{inv}$ = $1/C_{DOS}$ + $z_{inv}/k_{Si}$ [65]) is contributed again in series with $C_{ox}$ in $C_{gc}$, i.e., $1/C_{gc}$ = $1/C_{ox}$ + $1/C_{inv}$. Therefore, $C_{gc}$ changes with the gate bias from depletion to inversion [63,64].

Now consider how $C_{gc}$ can be expressed in the case of monolayer MoS$_2$ for which the thickness is negligible, because $C_{gc}$ in Si changes essentially as a function of the depletion layer thickness. For monolayer MoS$_2$, theoretical analysis suggests that the quantum capacitance ($C_Q$) for the monolayer channel contributes to $C_{gc}$ [66]. When the voltage is applied between the source and top gate, the top gate electrode and monolayer MoS$_2$ act as a parallel plate capacitor with a top gate insulator, where the same numbers of carriers with different polarities are induced. From the viewpoint of energy, the extra kinetic energy is required for monolayer MoS$_2$ to induce the same numbers of carriers in MoS$_2$ as those in metal because of the small DOS of monolayer MoS$_2$, that is unlike that of a metal. That is, because the extra energy in the system causes the voltage drop in the equivalent circuit, the additional voltage drop that is consumed in the MoS$_2$ channel should be considered. This is introduced as the capacitor, i.e., $C_Q$ = $e^2$DOS, rather than resistance R or inductor Z because the carriers are "accumulated" in the channel. Therefore, $C_{gc}$ can be expressed as $1/C_{gc}$ = $1/C_{ox}$ + $1/C_Q$. Moreover, the contribution from the interface states density ($C_{it}$ and $R_{it}$) is introduced in parallel to $C_Q$ in the equivalent circuit. Finally, the contributions from (A) parasitic capacitance and (B) channel resistance must be included in the equivalent circuit because the actual device is the FET structure, not the metal/oxide/semiconductor capacitor (MOSCAP). Then, the equivalent circuit examined experimentally is shown in **Fig. 7(a)** [17,18].

First, the validity of the equivalent circuit in **Fig. 7(a)** is verified by experimentally extracting $C_Q$ from the C-V measurement. The monolayer MoS$_2$ FET was fabricated on the SiO$_2$/Si substrate by mechanical exfoliation. The 10-nm Al$_2$O$_3$ top gate oxide with a 1-nm Y$_2$O$_3$ buffer layer was deposited via atomic layer deposition, followed by the Al top-gate electrode formation [67]. The analytical values for $C_Q$ at different temperatures are shown in **Fig. 7(c)**, where the shape of 2D DOS can be clearly observed due to $C_Q$ = $e^2$DOS and the temperature dependence results from the Fermi-Dirac distribution. Experimentally, because the channel area is much smaller than that between the high-doped Si substrate and the metal electrode pad, the large contribution of $C_{para}$ is an important problem. To solve this problem, an insulating quartz substrate was used. **Figure 7(b)** clearly indicates that $C_{para}$ can be neglected in $C_{gc}$ for the monolayer $MoS_2$ on the quartz substrate compared with that on the $SiO_2$/Si substrate. Because the issue of $C_{para}$ is clearly solved, the equivalent circuit is reconsidered. Generally, the trapping/detrapping of the carriers at the interface states cannot follow the high frequency of 1 MHz. Therefore, the terms of $C_{it}$ and $R_{it}$ can be neglected at 1 MHz. Moreover, in the top gate voltage ($V_{TG}$) range from near the threshold voltage ($V_{th}$) to the accumulation, $R_{ch}$ is low enough to be neglected, simplifying the equivalent circuit to $1/C_{gc} = 1/C_{ox} + 1/C_Q$, as shown in the inset of **Fig. 7(c)**. Because the capacitance of the top gate insulator was measured as $C_{ox}$ = 0.35 μFcm$^{-2}$ in **Fig. 7(b)**, it is possible to extract $C_Q$ at different temperatures, as shown in **Fig. 7(c)**. The experimental data are in good agreement with the theoretical solid lines.

Finally, the channel capacitances of Si-MOSFET and monolayer $MoS_2$ FET are compared. For Si, the inversion capacitance can be expressed using two terms, i.e., (i) DOS and (ii) distance $z_{inv}$ (**Fig. 2(a)**) as $1/C_{inv} = 1/C_{DOS} + z_{inv}/k_{Si}$ [65]. For the monolayer $MoS_2$, the thickness of the channel can be neglected. That is, only the $C_{DOS}$ term can be considered ($C_Q = C_{DOS}$) for monolayer $MoS_2$. It should be noted that $C_{ox}$ for the present top gate insulator (~0.35 μFcm$^{-2}$) is large enough to extract $C_Q$ (~80 μFcm$^{-2}$) of the monolayer $MoS_2$, unlike the conventional back gate $SiO_2$ with a thickness of 90 nm (~0.0383 μFcm$^{-2}$). As long as the $SiO_2$ back gate is used, $C_Q$ cannot be extracted. In other words, $C_Q$ can be neglected in our general analysis.

### 4.2 Depletion and inversion in multilayer MoS$_2$ FET

The channel capacitance for monolayer $MoS_2$ is determined by $C_Q$. Here, the naive idea "S vacancies dope $MoS_2$ crystal with electrons and act as fixed ions. Therefore, the depletion layer is formed under the positive gate bias, resulting in the off state" is experimentally verified based on the *C-V* measurements with increasing $MoS_2$ thickness ($t_{ch}$). **Figure 8** shows (a) the conductivities and (b) normalized $C_{gs}$ as a function of $t_{ch}$. It is observed that $I_{off}$ increased with increasing channel thickness. This can be understood from the schematics in **Fig. 8(c)**. For the monolayer case, it can be considered that $E_F$ can be modulated due to the small DOS, that is, $E_F$ can be moved within $E_g$, leading to the off state. For the multilayer $MoS_2$ case ($t_{inv} < t_{ch} < W_{Dm}$), where $W_{Dm}$ is the maximum depletion width, the depletion layer can reach the channel thickness by applying the negative gate bias, and a further negative bias results in the inversion layer formation. However, the access region of the $MoS_2$ channel is not modulated by the top gate and is still *n* type. Therefore, the *p-n* junction prevents the carrier transport, leading to the current off state. This is because the generation/recombination current can be neglected due to the large energy gap of ~2 eV for the multilayer. Moreover, even for the device where the top gate fully covers the entire channel, the carrier injection at the metal/$MoS_2$ contact is limited to the conduction band because of the strong Fermi-level pinning [62,68]. The *p*-type conduction is prevented, and the current off state is observed. Alternatively, from the viewpoint of *C-V* measurements on the multilayer $MoS_2$ ($t_{inv} < t_{ch} < W_{Dm}$), $C_D$ can be considered when a negative gate bias is applied. Once the depletion layer reaches the channel thickness, the electrical communication between the source and channel is limited to the negligibly small amounts of carriers in the depleted region where the $C_Q$ ascribed to the DOS of multilayer $MoS_2$ should be considered. In this case, $C_{gc}$ asymptotically approaches zero [18].

For bulk $MoS_2$ ($t_{ch} > W_{Dm}$), the bottom part of bulk $MoS_2$ is not depleted, and $MoS_2$ remains *n*-type because the channel thickness is greater than $W_{Dm}$. Therefore, the current off state is not achieved, as shown in **Fig. 8(a)**. By contrast, in the *C-V* measurements, the carriers extruded to the edge of the depletion layer can electrically communicate with the source, and the contribution of $C_D$ (=$k_{MoS2}/W_{Dm}$) can be clearly observed to be nonzero, as indicated in **Fig. 8(b)**. This is supported by the fact that $C_{para}$ can be fully neglected due to the use of the quartz substrate. From the *I-V* data, $W_{Dm}$ is estimated to be ~48-55 nm, and the donor density of ~2-3×10$^{17}$ cm$^{-3}$ is further calculated based on $W_{Dm}$. This donor density is in good agreement with the S vacancy density in bulk $MoS_2$.

Based on these considerations, it is found that the above-mentioned naive idea is basically reasonable. Moreover, the fact that $W_{Dm}$ is detected in the *C-V* curve of bulk $MoS_2$ provides experimental evidence of the inversion layer formation under the negative gate bias. Now extend the fact that "defects themselves act similarly to the substitutional doping in $MoS_2$" to all of the 2D system. Under this assumption, the defect densities for many kinds of 2D crystals can be calculated using $W_{Dm}$, which is estimated experimentally from the layer number dependence in *I-V* curves reported in the literature [18]. **Figure 9** summarizes $W_{Dm}$ as a function of the defect density. It should be noted that the defect densities are slightly overestimated for 2D crystals since the relationship between $W_{Dm}$ and $N_D$ ($N_A$) for $MoS_2$ with small intrinsic carrier concentration is used in this plot. From this figure, it can be understood that $WS_2$ possesses the lowest defect density, while many defects are introduced in $PtS_2$. Of course, although this figure can be revised by the improvement of the crystal growth technique, this figure can be used as the entire picture of crystallinity for many kinds of 2D crystals at present.

### 4.3 Origin of $D_{it}$ in *n*-type and *p*-type MoS$_2$ FET

The basic operation mechanism in $MoS_2$ FET was revealed above. Next, let us discuss how to improve the FET performance of MoS$_2$. From the viewpoint of both the ultralow power consumption and high-speed operation, a high-$k$ gate insulator is required. Although $h$-BN is highly compatible with 2D channels, as seen in **chapter 3**, its low dielectric constant ($k$ = ~3) is unfavorable. To extract the potential of high-$k$ electrostatics due to the atomic thickness of 2D channel, understanding of the high-$k$/monolayer MoS$_2$ interface is quite important for practical use. Although some studies of the interface properties have been reported [6-14], common understanding of the origin of the interface degradation has not been obtained. To date, several physical origins for $D_{it}$ have been proposed, namely (I) the defect states induced by V$_S$, (II) traps in the high-$k$ insulator, and (III) the strain in MoS$_2$ induced externally, as summarized in **Fig. 10(a)**.

Although the details are not described here, (II) traps in the high-$k$ insulator are not the dominant origin. The electrical quality of the high-$k$ gate insulator on 2D channels is indeed not significant compared with that on the Si substrate. Nevertheless, the trap sites induced in the 2D channel itself are more dominant at the present stage because the atomically thin channel is quite sensitive to the surroundings. Therefore, the influence of the traps in the high-$k$ gate insulator should be realized after the trap sites in the 2D channels are eliminated. Two types of origins, (I) and (III), are focused on here. **Figure 10(b)** shows a schematic of the defect states for V$_S$ in monolayer MoS$_2$ for origin (I) as evaluated by density functional theory (DFT) [48]. The defect states for V$_S$ are formed near the mid-gap and valence band top. To reveal the origin of $D_{it}$ from the comparison with the DFT calculations, the energy distribution of $D_{it}$ should be obtained experimentally. However, a recent study [18] indicated that the conventional $C$-$V$ method for the $D_{it}$-energy relation developed for Si systems cannot be simply applied to the FET structure of 2D channels because the channel charging process due to the high channel resistance at the depletion is more dominant than the electron capture/emission process by the interface traps [28]. Therefore, to obtain the energy distribution of $D_{it}$, the modeling of $I_D$-$V_{TG}$ characteristics [17,18] was performed by considering the MoS$_2$ channel carrier statistics through $C_Q$ and its transport through the Drude model [21]. Moreover, as discussed in **section 2.2**, the conduction and valence bands of MoS$_2$ are formed by the $d$-$d$ splitting in Mo due to the interaction with the ligand field of S. Therefore, this electronic band structure is expected to be strongly affected by the strain in MoS$_2$, which is related to origin (III). To remove the strain caused by the surface roughness of the substrate, atomically flat $h$-BN is used as the substrate. Here, the various gate stack structures, as shown in **Fig. 11(a)**, are compared to separate two different origins, (I) and (III). Then, the dominant origin of $D_{it}$ for $n$-type and $p$-type MoS$_2$ is discussed.

**Figure 11** shows the transfer characteristics for (b) the $n$-type monolayer MoS$_2$ and for (c) the $p$-type four-layer MoS$_2$. The energy distributions of $D_{it}$ extracted from the transfer characteristics are summarized in **Fig. 12**. First, the $n$-type MoS$_2$ is discussed. For the high-$k$/MoS$_2$ interface in the structure (i), the S.S. value of 230 meV/dec is relatively high because the V$_S$ defects, as well as the strain in MoS$_2$ due to the substrate surface roughness and high-$k$ deposition, are expected. By contrast, for the MoS$_2$/$h$-BN structure (ii), where the strain in MoS$_2$ is excluded due to the absence of high-$k$ deposition and the atomically flat surface of the substrate, nearly ideal transfer characteristics with a low S.S. value of 75 meV/dec and negligible hysteresis are observed. $D_{it}$ for the MoS$_2$/$h$-BN structure is considerably reduced from that for the high-$k$/MoS$_2$ structure, as shown in **Fig. 12**. Because the MoS$_2$ monolayers for both devices are prepared from the same MoS$_2$ crystals by mechanical exfoliation, the defect density for V$_S$ should be the same. Moreover, the defect states for V$_S$ near the mid-gap level are somewhat far from the extracted $D_{it}$ range. Therefore, the origin of $D_{it}$ at the conduction band side may be the bond bending of the Mo $d$ orbital due to the external strain (III), rather than the defect state induced by V$_S$ (I). Further analysis suggests that the strain due to high-$k$ deposition is more dominant than that due to the surface roughness of the substrate. A detailed description of these experiments is found in ref. [21]. In the above discussion, it was concluded that defects states of V$_S$ are not related to S.S. However, the reduction in the S vacancy concentration is clearly important. Indeed, the S vacancies on both top and bottom surfaces were repaired by the thiol chemical route from $8.1 \times 10^{12}$ cm$^{-2}$ to $5.22 \times 10^{12}$ cm$^{-2}$, resulting in the high mobility of ~80 cm$^2$/Vs at room temperature, that is, the high on-current [69].

Next, let us consider the $p$-type MoS$_2$ that was prepared by Nb doping ($N_A$ = ~2×10$^{19}$ cm$^{-3}$) during the synthesis [70] since the substitution of the Mo site by Nb (Nb$_{Mo}$) is thermodynamically stable [71]. It is quite interesting that $D_{it}$ was not improved at all, even after the heterostructure formation with $h$-BN (structure (v)), suggesting that the strain (II) is not the dominant origin for the valence band side. As shown in **Fig. 10(b)**, the defects states for Nb$_{Mo}$ are formed just below the valence band maximum [71]. However, interestingly, the defect density of Nb$_{Mo}$ is smaller than that of V$_S$, even in the Nb-doped $p$-type MoS$_2$ crystal [21,72]. Moreover, it is reported that $p$-type conduction for the "$n$-type" monolayer MoS$_2$ FET was not detected, even using the ion gating; instead, the abnormal conductance peak was observed due to the defect states of V$_S$ [73]. These results suggest that the defect density of V$_S$ near the valence band is much larger than that near the mid-gap. Therefore, for the valence band side, the defect states of $V_S$ may be the dominant origin.

Based on the above discussion, to achieve the sharp S.S. in the CMOS operation using 2D channels, the crystalline quality of *p*-type $MoS_2$ should be improved. To date, the substitutional doping technique using conventional ion-implantation has not been established yet. Because the *p*-type 2D crystals except Nb-doped $MoS_2$ are quite limited, investigation of other *p*-type 2D crystals is urgently needed. Moreover, improvement of the high-*k*/2D interface is still an important issue. As discussed above for *n*-type $MoS_2$, since the defect states in $E_g$ are not related to the FET operation, the reduction of the external strain may be the dominant issue. Therefore, the top gate $MoS_2$ FET with monolayer *h*-BN buffer layer has been reported [74]. Other buffer layer materials except *h*-BN should be explored because *h*-BN has a low dielectric constant, and its inactive surface prevents it from acting as the nucleation site for ALD. Recently, the 3,4,9,10-perylene-tetracarboxylic dianhydride (PTCDA) monolayer organic layer was used as the ALD-active buffer layer, and an effective oxide thickness (EOT) of 1 nm and S.S. of 60 meV/dec was achieved by ALD for $HfO_2$ [75]. Because the realization of an atomically flat surface by the monolayer organic layer has been demonstrated, the utilization of an organic buffer layer may become one of the future design guidelines for the gate stack.

## 5. Summary & Outlook

In this review, the interface properties specific to 2D layered channels are described by comparison to the properties of the $SiO_2$/Si interface. The significant advantage of 2D materials is their large surface/volume ratio, enabling their use for the fabrication of highly sensitive sensors. In other words, it is quite difficult to control the 2D channels from the viewpoint of the miniaturized FETs in the integrated circuit because they are so sensitive to the fluctuation of their surroundings. This problem is partially addressed in the present review. A paradigm shift may be required to overcome this issue. In the conventional Si processes, all of the building blocks of the devices are composed of inorganic materials. It is possible to treat graphene with its strong $sp^2$ bonding as a conventional inorganic semiconductor. On the other hand, for other 2D materials, the melting points are often low despite the covalent bonding due to the ligand field. This suggests that the thermal budget is quite low. Therefore, it may be important to recognize that the 2D channels are equivalent to organic semiconductors. The inclusion of organic materials in 2D device structures may lead to a breakthrough in the improvement of the 2D device performance. In this case, reliability will be the key issue for practical applications.

## Acknowledgements

The research described in this review was conducted together with our lab members and Drs. T. Taniguchi and K. Watanabe, (NIMS), Prof. K. Ueno (Saitama Univ.), and Prof. X. Wang (Nanjin Univ.). We truly appreciate all of the discussions. We are grateful to Covalent Materials for kindly providing us with Kish graphite. This research was supported by The Canon Foundation, the JSPS Core-to-Core Program, A. Advanced Research Networks, the JSPS A3 Foresight Program, and JSPS KAKENHI Grant Numbers JP16H04343, JP19H00755, and 19K21956, Japan.

**References**
[1] Desai S B, Madhvapathy S R, Sachid A B, Llinas J P, Wang Q, Ahn G H, Pitner G, Kim M J, Bokor J, Hu C, Wong H -S P and Javey A 2016 *Science* **354** 99
[2] Sarkar D, Xie X, Liu W, Kang J, Gong Y, Kraemer S, Ayajan P M and Banerjee K 2015 *Nature* **526** 91
[3] Akinwande D, Huyghebaert C, Wang C -H, Serna M I, Goossens S, Li L -J, Wong H -S P and Koppens F H L 2019 *Nature* **573** 507
[4] Kang K, Xie S, Hunag L, Han Y, Huang P Y, Mak K F, Kim C -J, Muller D and Park J 2015 *Nature* **520** 656
[5] Yu H, Liao M, Zhao W, Liu G, Zhou X J, Wei Z, Xu X, Liu K, Hu Z, Deng K, Zhou S, Shi J -A, Gu L, Shen C, Zhang T, Du L, Xie L, Zhu J, Chen W, Yang R, Shi D and Zhang G 2017 *ACS nano* **11** 12001
[6] Chen X, Wu Z, Xu S, Wang L, Huang R, Han Y, Ye W, Xiong W, Han T, Long G, Wang Y, He Y, Cai Y, Sheng P and Wang N 2015 *Nature Commun.* **6** 6088
[7] Zou X, Wang J, Chiu C H, Wu Y, Xiao X, Jiang C, We W, Mai L, Chen T, Li J, Ho J C and Liao L 2014 *Adv. Mater.* **26** 6255
[8] Zhu W, Low T, Lee Y -H, Farmar D B, Kong J, Xia F and Avouris P 2014 *Nature Commun.* **5** 3087
[9] Ninomiya N, Mori T, Uchida N, Watanabe E, Tsuya D, Moriyama S, Tanaka M and Ando A 2015 *Jpn. J. Appl. Phys.* **54** 046502
[10] Choi K, Raza S R A, Lee H S, Jeon P J, Pezeshki A, Min S W, Kim J S, Yoon W, Ju S, Lee K and Im S 2015 *Nanoscale* **7** 5617
[11] Takenaka M, Ozawa Y, Han J and Takagi S 2016 *IEEE International Electron Devices Meeting Tech. Dig.* 139
[12] Park S, Kim S Y, Choi Y, Kim M, Shin H, Kim J and Choi W 2016 *ACS Appl. Mater. Interfaces* **8** 11189
[13] Vu Q A, Fan S, Lee S H, Joo M -K, Yu W J andLee Y H 2018 *2D Mater.* **5** 031001
[14] Zhao P, Khosravi A, Azcatl A, Bolshakov P, Mirabelli G, Caruso E, Hinkle C, Hurley P K, Wallace R M and Young C D 2018 *2D Mater.* **5** 031002
[15] Kanayama K and Nagashio K 2015 *Sic. Rep.* **5** 15789
[16] Uwanno T, Taniguchi T, Watanabe K and Nagashio K 2018 *ACS appl. mater. interfaces* **10** 28780
[17] Fang N and Nagashio K *J. Phys. D* **51** 065110

[18] Fang N and Nagashio K 2018 *ACS Appl. Mater. interfaces* **10** 32355
[19] Taniguchi K, Fang N and Nagashio K 2018 *Appl. Phys. Lett.* **113** 133505
[20] Fang N and Nagashio K *2D Mater.* **7** 014001
[21] Fang N, Toyoda S, Taniguchi T, Watanabe K and Nagashio K 2019 *Adv. Func. Mater.* **29** 1904465
[22] Hisamoto D, Lee W -C, Kedzierski J, Takeuchi H, Asano K, Kuo C, Anderson E, King T -J, Bokor J and Hu C 2000 *IEEE Trans. Electron Dev.* **47** 2320
[23] Leong M, Doris B, Kedzierski J, Rim K and Yang M 2004 Science **306** 2057
[24] Yeric G 2015 *IEEE International Electron Devices Meeting Tech. Dig.* 1
[25] An T Y, Choe K K, Kwon K -W and Kim S Y 2014 *J. Semicond. Technol. Sci.* **14** 525
[26] Uchida K, Koga J, Ohba R, Numata T and Takagi S 2001 *IEEE International Electron Devices Meeting Tech. Dig.* 633
[27] Takagi S, Koga J, Toriumi A 1998 *Jpn. J. Appl. Phys.* **37** 1289
[28] Schmidt M, Lemme M C, Gottlob H D B, Driussi F, Selmi L and Kurz H 2009 *Solid-State Electronics* **53** 1246
[29] Cao W, Kang J, Sarkar D, Liu W and Banerje K 2015 *IEEE Trans. on Electron Devices* **62** 3459
[30] Li L, Yu Y, Ye G J, Ge Q, Ou X, Wu H, Feng D, Hui X, Chen H and Zhang Y 2014 *Nature Nanotechnol.* **9** 372
[31] Fang H, Chuang S, Chang T C, Takei K, Takahashi T and Javey A 2012 *Nano Lett.* **12** 3788
[32] Sugita Y, Watanabe S, Awaji N and Komiya S 1997 *Appl. Surf. Sci.* **100/101** 268
[33] Kosowsky S D, Pershan P S, Krisch K S, Bevk J, Green M L, Brasen D, Feldman L C and Rpoy P K *Appl. Phys. Lett.* **70** 3119
[34] Miyazaki S, Nishimura H, Fukuda M, Ley L and Ristein J 1997 *Appl. Surf. Sci.* **114** 585
[35] Oh J H, Yeom H W, Hagimoto Y, Ono K Oshima M, Hirashita N, Nywa M, and Toriumi A. and Kakizaki A. 2001 *Phys. Rev. B* **63** 205310
[36] Bongiorno A, Pasquarello A, Hybertsen M S and Feldoman L C *Phys. Rev. Lett.* **90** 186101
[37] Poindexter E H 1989 *Semicond. Sci. Technol.* **4** 961
[38] Sakurai T and Sugano T 1981 *J. Appl. Phys.* **52** 2889
[39] Castro Neto A H, Guinea F, Peres N M R, Novoselov K S and Geim A K 2009 *Rev. Modern Phys.* **81** 109
[40] Lee C, Wei X, Kysar J W and Hone J 2008 *Science* **321** 385
[41] Martin J, Akerman N, Ulbricht G, Lohmann T, Smet J H, von Klitzing K and Yacoby A 2008 *Nature Phys.* **4** 144
[42] Rhodes D, Chae S H, Ribeiro-Palau R, Hone J 2019 *Nature Mater.* **18** 541
[43] Wilson J A and Yoffe A D *Adv. Phys.* **18** 193
[44] Kolobov A V and Tominaga J 2016 *Two-Dimensional Transition-Metal Dichalcogenides* (Springer).
[45] Shin B G, Han G H, Yun S J, Oh H M, Bae J J, Song Y J, Park C -Y andLee Y H 2016 *Adv. Mater.* **28** 9378
[46] Watanbe K, Taniguchi T and Kanda H 2004 *Nature Mater.* **3** 404
[47] Wong D, Velasco Jr J, Ju L, Lee J, Kahn S, Tsai H -Z, Germany C, Taniguchi T, Watanabe K, Zettl A, Wnag F and Crommie M F 2015 *Nature Nanotechnol.* **10** 949
[48] Zhou W, Zou X, Najmaei S, Liu Z, Shi Y, Kong J and Idrobo J C 2013 *Nano Lett.* **13** 2615
[49] Hong J, Hu Z, Probert M, Li K, Lv D, Yang X and Zhang J 2015 *Nature Commun.* **6** 6293
[50] Xue J, Sanchez-Yamagishi J, Bulmash D, Jacquod P, Deshpande A, Watanabe K, Taniguchi T, Jarillo-Herrero P and BeRoy B J 2011 *Nature Mater.* **10** 282
[51] Oostinga J B, Heersche H B, Liu X, Morpurgo A F and Vandersypen M K 2008 *Nature Mater.* **7** 151
[52] Koshino M 2009 *New J. Phys.* **11** 095010
[53] Zhang Y, Tang T -T, Girit C, Hao Z, Martin M C, Zettl A, Crommie M F, Shen Y R and Wang F 2009 *Nature* **459** 820
[54] Miyazaki H, Tsukagoshi K, Kanda A, Otani M and Okada S 2010 *Nano Lett.* **10** 3888
[55] Takahashi N and Nagashio K 2016 *Appl. Phys. Express* **9** 125101
[56] Kanayama K, Nagashio K, Nishimura T and Toriumi A 2014 *Appl. Phys. Lett.* **104** 083519
[57] Uwanno T, Hattori Y, Taniguchi T, Watanabe K and Nagashio K 2015 *2D Mater.* **2** 041002
[58] Toyoda S, Uwanno T. Taniguchi T. Watanabe K and Nagashio K 2019 *Appl. Phys. Express* **12** 055008
[59] Beidenkopf H, Roushan P, Seo J, Gorman L, Drozdov I, Hor Y S, Cava R J and Yazdani A 2011 *Nature Phys.* **7** 939
[60] Cao Y, Fatemi V, Fang S, Watanabe K, Taniguchi T, Kaxiras E, and Jarillo-Herrero P 2018 *Nature* **556** 43
[61] Osada M and Sasaki T 2012 *Adv. Mater.* **24** 210
[62] Das S, Chen H-Y, Penumatcha A V and Appenzeller J 2013 *Nano lett.* **13** 100.
[63] Sze S M and Ng K K 2007 *Physics of Semiconductor Devices*, Wiley-interscience, NJ
[64] Nicollian E H and Brews J R 1982 *MOS Physics and Technology*, Wiley, NY
[65] Nakajima Y, Horiguchi S, Shoji M and Omura Y 1998 *J. Appl. Phys.* **83** 4788
[66] Ma N and Jena D 2015 *2D Materials* **2** 015003
[67] Kurabayashi S and Nagashio K 2017 *Nanoscale* **9** 13264
[68] Kim C, Moon I, Lee D, Choi M S, Ahmed F, Nam S, Cho Y, Shin H -J, Park S and Yoo W J 2017 *ACS nano* **11** 1588
[69] Yu Z, Pan Y, Shen Y, Wang Z, Ong Z-Y, Xu T, Xin R,

Pan L, Wang B, Sun L, Wang J, Zhang G, Zhang Y W, Shi Y, and Wang X 2014 *Nature commun.* **5** 5290

[70] Suh J, Park T E, Lin D Y, Fu D, Park J, Jung H J, Chen Y, Ko C, Jang C, Sun Y, Sinclair R, Chang J, Tongay S and Wu J 2014 *Nano Lett.* **14** 6976

[71] Dolui K, Rungger I, Pemmaraju C D andSanvito S 2013 *Phys. Rev. B* **88** 075420

[72] Siao M D, Shen W C, Chen R S, Chang Z W, Shih M C, Chiu Y P, Cheng C M 2018 *Nature Commun.* **9** 1442

[73] Ponomarev E, Pasztor A, Waelchli A, Scarfato A, Ubrig N, Renner C, Morpurgo A F 2018 *ACS Nano* **12** 2669

[74] Zou X, Huang C -W, Wang L, Yin L -J, Li W, Wang J, Wu B, Liu Y, Yao Q, Jiang C, Wu W -W, He L, Chen S, Ho J C and Liao L 2016 *Adv. Mater.* **28** 2062

[75] Li W, Zhou J, Cai S, Yu Z, Zhang J, Fang N, Li T, Wu Y, Chen T, Xie X, Ma H, Yan K, Dai N, Wu X, Zhao H, Wang Z, He D, Pan L, Shi Y, Wang P, Chen W, Nagashio K, Duan X and Wang X 2019 *Nature Electronics* **2** 563

[76] Vancsó P, Magda G Z, Pető J, Noh J Y, Kim Y S, Hwang C, Biró L P and Tapasztó L 2016 *Sci. Rep.* **6** 29726

**Figures**

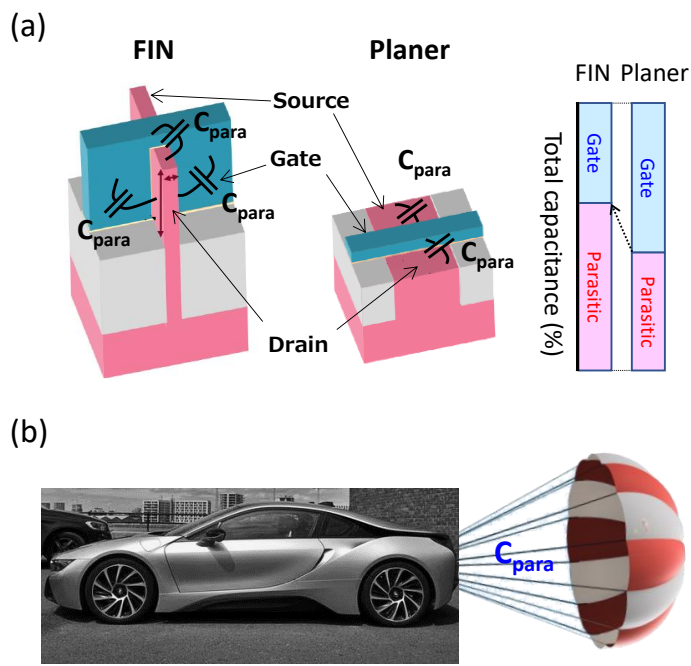

**Fig. 1** Schematics of FinFET and planar FET. A high-performance race car with a large parachute is a metaphor for FinFET [24].

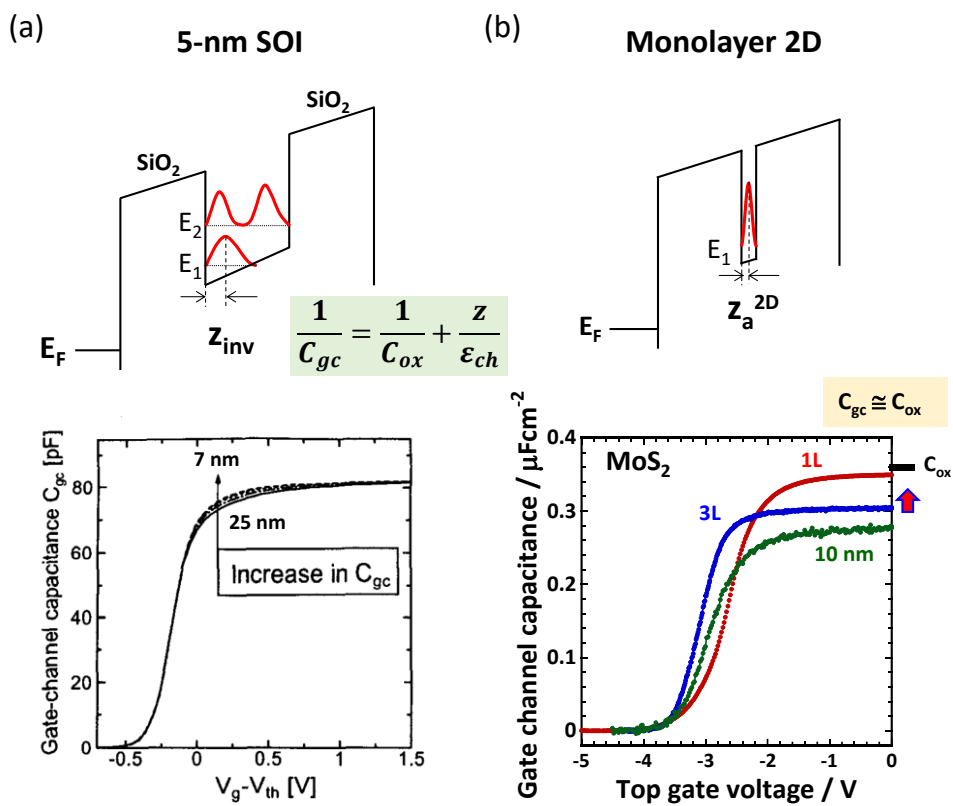

**Fig. 2** Electron density distribution and $C_{gc}$ as a function of gate voltage for (a) 5-nm SOI [26] and (b) monolayer MoS$_2$ [20].

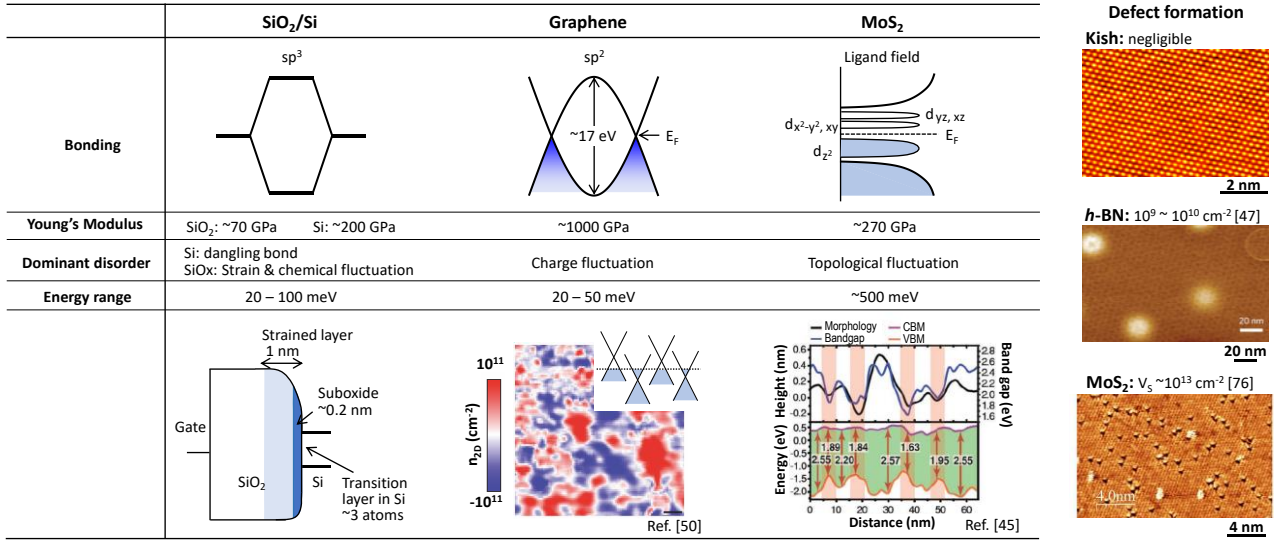

**Fig. 3** Comparison of SiO$_2$/Si, graphene and MoS$_2$.

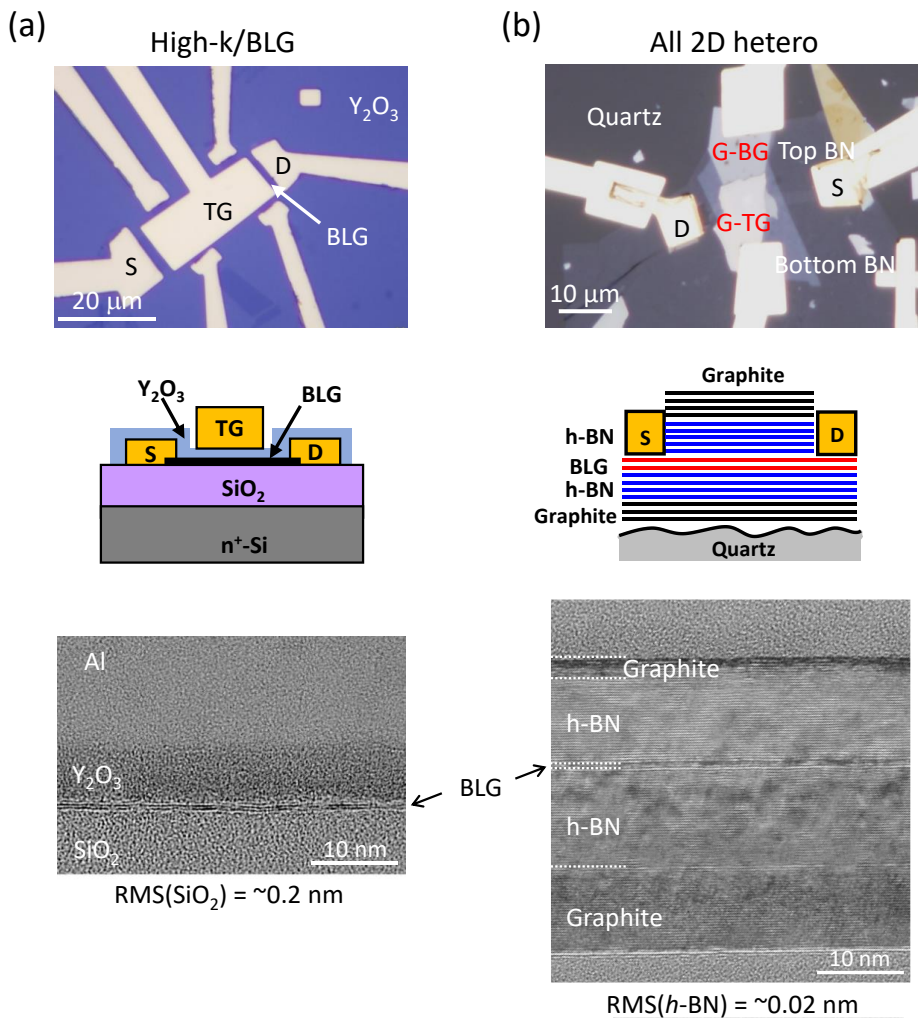

**Fig. 4** Device structure and cross-sectional TEM image for (a) high-$k$/BLG gate stack and (b) all 2D hetero.

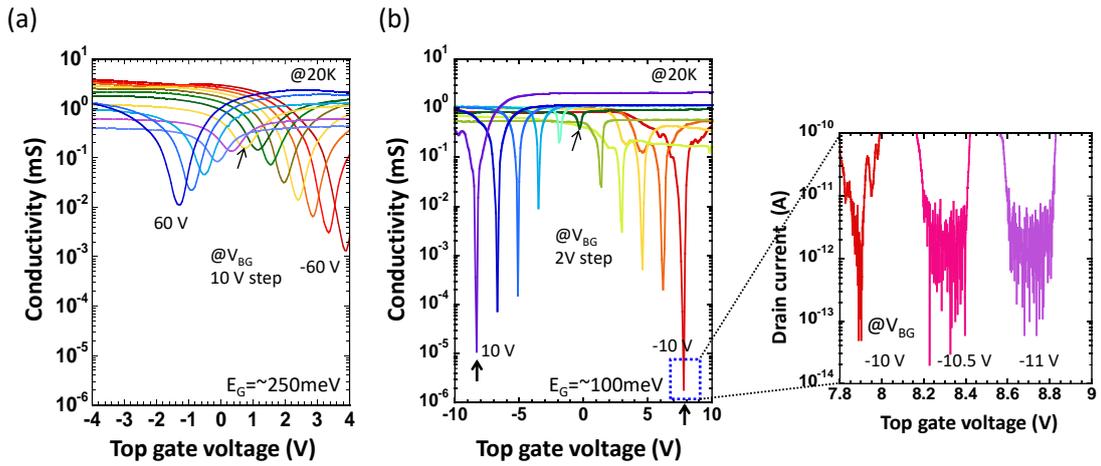

**Fig. 5** Two-terminal conductivity as a function of top gate voltage at different back gate voltage at 20 K for (a) high-*k*/BLG gate stack and (b) all 2D hetero.

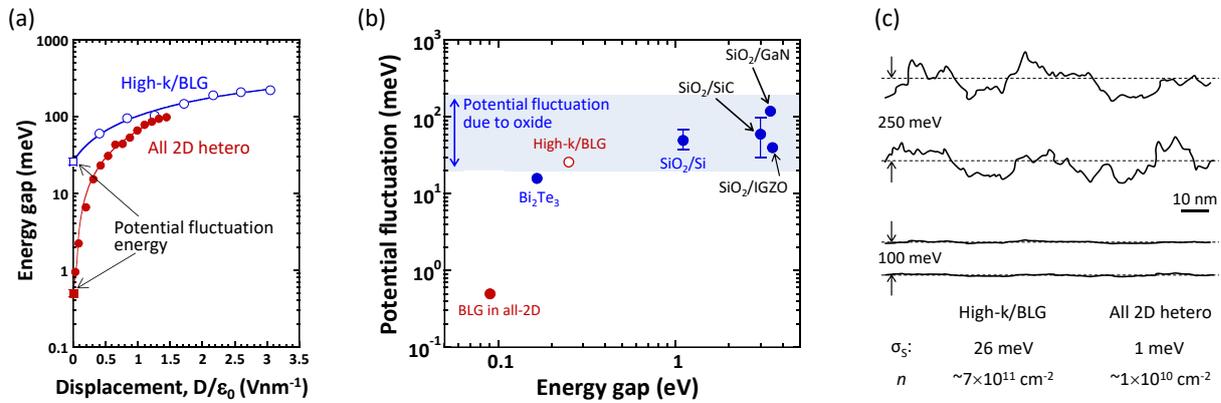

**Fig. 6** (a) $E_g$ as a function of $\bar{D}$ for high-*k*/BLG gate stack and all 2D hetero. (b) Comparison of potential fluctuations in various insulator/semiconductor gate stack structures. The references for the data shown in this figure can be found in Ref. [16]. (c) Schematics of the spatial band gap fluctuations for high-*k*/BLG gate stack and all 2D hetero.

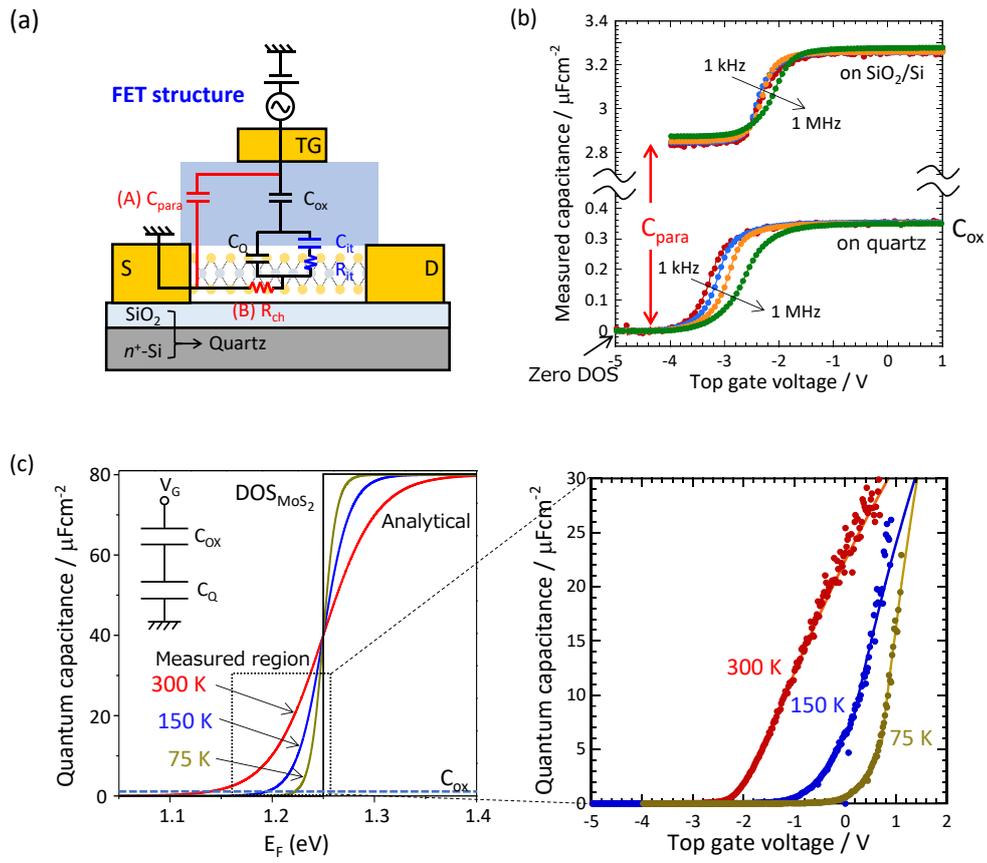

**Fig. 7** (a) Equivalent circuit for top gate MoS$_2$ FET. (b) Cgc as a function of the top gate voltage for monolayer MoS$_2$ FET on the SiO$_2$/Si substrate and the quartz substrate. (c) $C_Q$ as a function of EF at different temperatures. The inset shows the simplified equivalent circuit. The left figure is the theoretical calculation, while the right figure shows the experimental data.

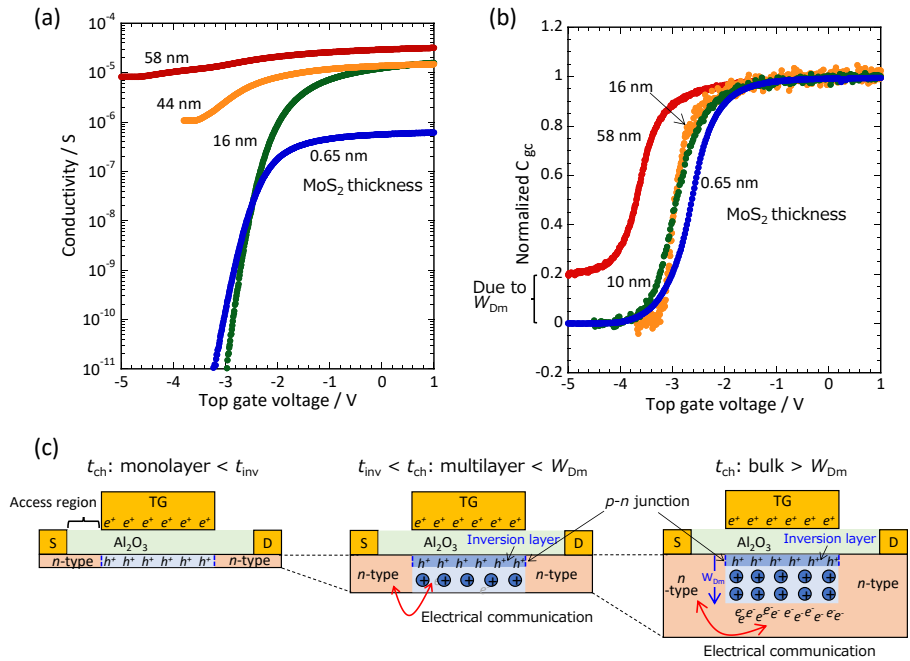

**Fig. 8** (a) Conductivity as a function of top gate voltage for different channel thicknesses of MoS$_2$. (b) Normalized Cgc as a function of top gate voltage for different channel thicknesses of MoS$_2$. (c) Schematics of MoS$_2$ FET operation.

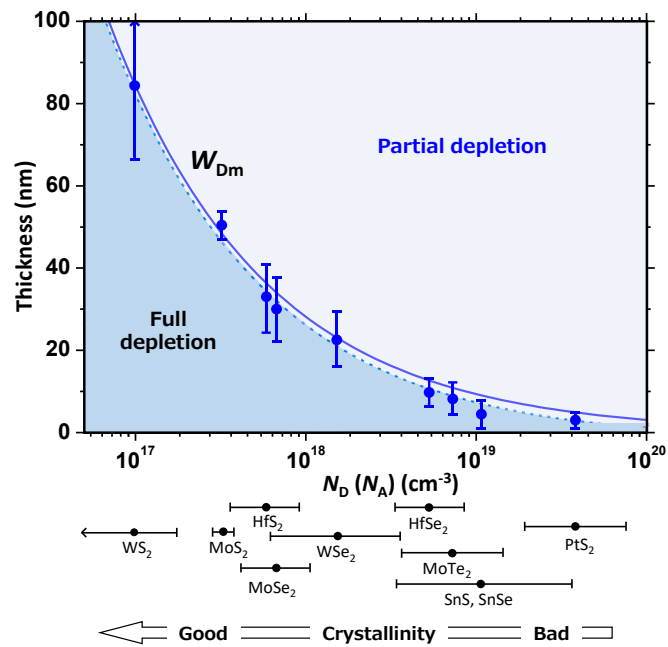

**Fig. 9** $W_{Dm}$ as a function of $N_D$ ($N_A$). The references for the data shown in this figure can be found in Ref. [18].

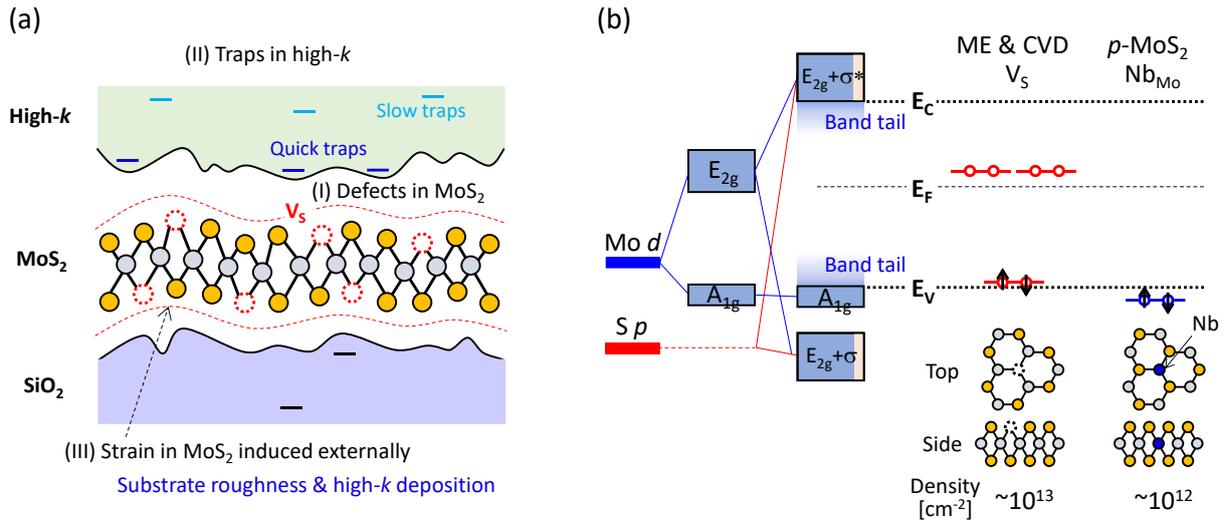

**Fig. 10** (a) Schematic illustration of different origins of interface states in high-$k$/MoS$_2$/oxide structures. (b) Schematic bonding diagram showing how the local orbitals on Mo and S interact to form CB, VB, and interface states in MoS$_2$. V$_S$ and Nb$_{Mo}$ represent the sulfur vacancy and the substitution of Mo site by Nb, respectively.

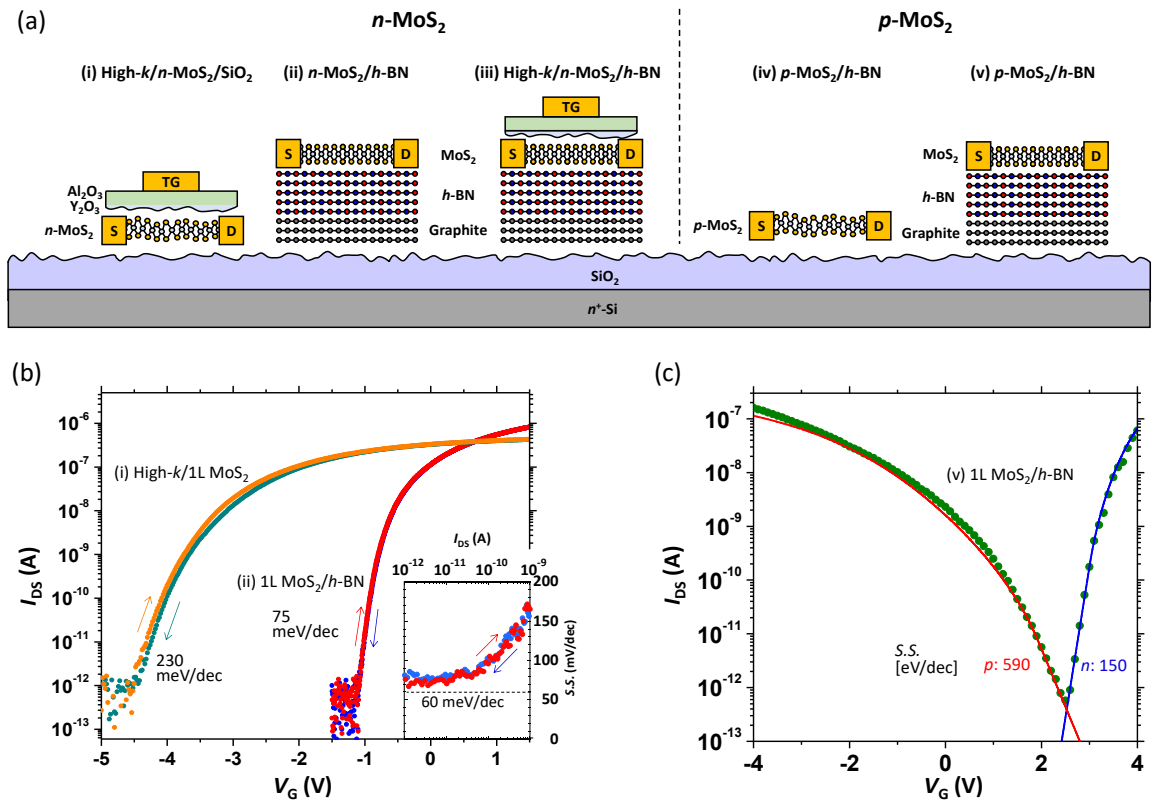

**Fig. 11** (a) Schematics of different device structures for $n$-type and $p$-type MoS$_2$ examined in the experiment. Because some of these are not described in this review, please refer [21]. (b) Transfer characteristics for $n$-type monolayer MoS$_2$ with device structure of (i) and (ii). The inset shows the S.S. value for (ii). (b) Transfer characteristics for $p$-type four-layer MoS$_2$ with device structure (v).

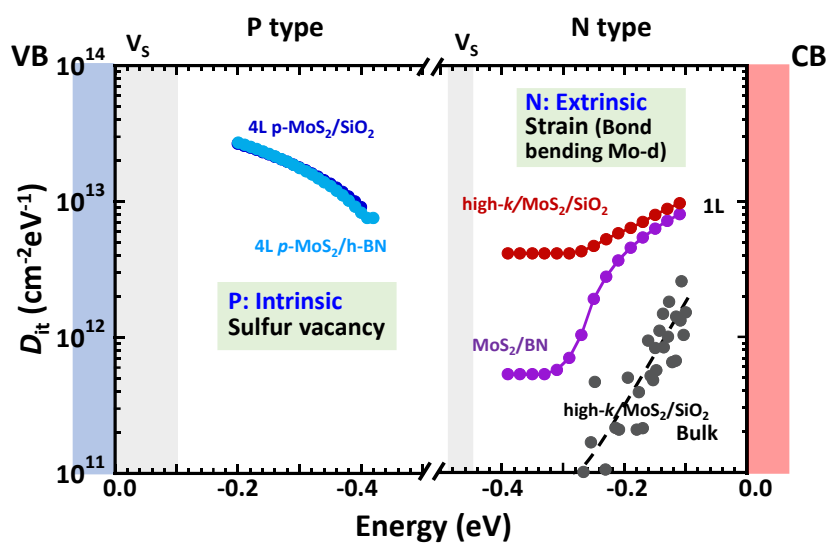

**Fig. 12** Full energy spectra of $D_{it}$ for different gate stack structures. Notice that the band gaps of 1L, 4L, and bulk MoS$_2$ are different. Therefore, the transverse axis for the $D_{it}$-energy distribution is shown as the energy from the CB/VB edge.